\documentclass[12pt]{article}
\usepackage{latexsym,epsfig,amssymb,amsmath}
\usepackage{slashed}

\newcommand{\la}{\label}

\newcommand{\eq}[1]{\begin{equation}#1\end{equation}}

\newcommand{\spl}[1]{\begin{split}#1\end{split}}

\newcommand{\boxedeq}[1]{
\begin{equation}
\fbox{
\rule[0.7cm]{0pt}{0pt}
$#1$
\rule[-0.45cm]{0pt}{0pt}
}
\end{equation}
}

\newcommand{\ft}[2]{{\textstyle\frac{#1}{#2}}}

\def\bfone{\relax{\rm 1\kern-.35em 1}}

\newcommand{\be}{\begin{equation}}
\newcommand{\ee}{\end{equation}}
\newcommand{\ben}{\begin{displaymath}}
\newcommand{\een}{\end{displaymath}}
\newcommand{\bea}{\begin{eqnarray}}
\newcommand{\eea}{\end{eqnarray}}
\newcommand{\nn}{\nonumber}

\newcommand{\bean}{\begin{eqnarray*}}
\newcommand{\eean}{\end{eqnarray*}}
\newcommand{\beqs}{\begin{eqnarray}}
\newcommand{\eeqs}{\end{eqnarray}}
\newcommand{\beal}{\begin{align}}

\newcommand{\mathon}{\mathversion{bold}}
\newcommand{\mathoff}{\mathversion{normal}}

\makeatletter
\@addtoreset{equation}{section}
\makeatother

\begin{document}

\thispagestyle{empty}

\begin{flushright}\small 
MIFPA-12-43 
\end{flushright}

\mathon
\vskip 10mm
\begin{center}
  {\LARGE {\bf Rigid $6D$ supersymmetry\\[.7ex] and localization}}
\end{center}
\mathoff


\bigskip

\begin{center}
{\bf Henning Samtleben$^a$, Ergin Sezgin$^b$, Dimitrios Tsimpis$^c$}\\[3ex]

$^a$\,{\em Universit\'e de Lyon, Laboratoire de Physique, UMR 5672, CNRS et ENS de Lyon,\\
46 all\'ee d'Italie, F-69364 Lyon CEDEX 07, France} \\
\vskip 4mm

$^b$\,{\em George P. and Cynthia W. Mitchell Institute \\for Fundamental
Physics and Astronomy \\
Texas A\&M University, College Station, TX 77843-4242, USA}\\
\vskip 4mm

$^c$\,{\em Universit\'{e} de Lyon\\
UMR 5822, CNRS/IN2P3, Institut de Physique Nucl\'{e}aire de Lyon\\ 
4 rue Enrico Fermi,  
F-69622 Villeurbanne Cedex, France}\\
\end{center}

\vskip1.8cm
\begin{center} {\bf Abstract } \end{center}
\begin{quotation}\noindent
We construct rigid supersymmetric theories for interacting
vector and tensor multiplets on six-dimensional Riemannian spin manifolds. 
Analyzing the Killing spinor equations, we  
derive the constraints on these theories.
To this end, we reformulate the conditions for supersymmetry as a set of necessary and sufficient conditions on the geometry. 
The formalism is illustrated 
with a number of examples, 
 including manifolds that are
hermitian, strong K\"{a}hler with torsion. 
As an application, we show that the path integral of pure super Yang-Mills theory defined on a Calabi-Yau threefold ${\cal M}_6$ localizes on stable holomorphic bundles over ${\cal M}_6$.

\end{quotation}

\vfill

December 2012

\newpage
\setcounter{page}{1}

\tableofcontents



\section{Introduction}


Recently, supersymmetric field theories on curved backgrounds have regained considerable
attention, in particular for the application of localization techniques in the computation of indices, 
partition functions and Wilson loops, see 
e.g.~\cite{Romelsberger:2005eg,Pestun:2007rz,Kapustin:2009kz,Okuda:2010ke,Hama:2010av}.
A systematic approach to the study of rigid supersymmetry in four-dimensional
curved space has been initiated in~\cite{Festuccia:2011ws} and further
developed in~\cite{Jia:2011hw,Samtleben:2012gy,Dumitrescu:2012ha,Liu:2012bi,Dumitrescu:2012at}.
Holographic applications of these theories have been studied in~\cite{Klare:2012gn,Cassani:2012ri} by embedding the curved space at the boundary of an asymptotically AdS space.
A similar analysis of supersymmetric theories 
has been performed for curved five-dimensional spaces in~\cite{Hosomichi:2012ek,Terashima:2012ra}  and in~\cite{Closset} for three-manifolds.
The construction of these curved theories is based on the existence of an underlying off-shell supergravity
in which the full off-shell supergravity multiplet is set to classical background values. The consistency
of this limit requires the existence of solutions of the
corresponding Killing spinor equations which in turn poses non-trivial constraints 
on the background fields.\footnote{
A recent approach for the construction of theories starting from
an on-shell formulation of supergravity has been discussed in~\cite{Kehagias:2012fh}.}

In the present paper we will focus on rigid supersymmetric theories in six-dimensional Riemannian spin manifolds. We will study the coupling of off-shell Yang-Mills (YM) multiplets to  a number of uncharged on-shell tensor multiplets. Such couplings have been constructed in $6D$ flat space-time of Minkowski signature~\cite{Bergshoeff:1996qm}. In view of the applications mentioned above, we shall extend this model to curved background and in Euclidean signature.  One way to construct these theories is to start from the Euclidean version of off-shell supergravity coupled to YM and tensor multiplets, and then take the rigid limit in which the fields in the supergravity multiplet are frozen in a manner consistent with supersymmetry.
In particular, the vanishing of the fermionic fields in this multiplet would yield the Killing spinor equations.  However, while the $N=(1,0)$ off-shell supergravity theory in $6D$ with Minkowskian signature is well known~\cite{Bergshoeff:1985mz,Coomans:2011ih}, thanks to the superconformal tensor calculus methods~\cite{Bergshoeff:1985mz}, the general couplings to Yang-Mills and multi-tensor multiplets is still missing. Extra complications may arise in the passage to Euclidean signature. This suggests an alternative approach in which we begin by formulating the  Euclidean version of the Killing spinor equations with suitable background fields. 
We distinguish two cases depending on whether these equations descend 
from an off-shell theory with ($\theta\not=0$) or without ($\theta=0$) R-symmetry gauging.
In the former case, additional Killing spinor equations have to be imposed.
Next, we translate the model of~\cite{Bergshoeff:1996qm} to Euclidean signature and then elevate the supersymmetry parameter to be Killing.  Performing a Noether procedure which takes this into account then produces the full dependence of the couplings on the nontrivial background fields.
Following this approach we construct the non-trivial supersymmetric 
tensor/vector couplings on a curved background.

Next, we systematically
analyse the Killing spinor equations in six-dimensional Riemannian spin manifolds  ${\cal M}_6$ using $G$-structures and we derive the constraints for the existence of non-trivial solutions. We reformulate these constraints as a set of conditions on the geometry. The necessary and sufficient conditions for minimal supersymmetry 
are given in equations (\ref{tc}), (\ref{tc0}) below as constraints on an $SU(3)$ structure. In particular for 
a theory whose background descends from a supergravity with R-symmetry gauging ($\theta\not=0$),
these constraints require ${\cal M}_6$ to be hermitian, strong K\"{a}hler with torsion, but not necessarily conformally balanced. 
For $\theta=0$ on the other hand,
the manifold ${\cal M}_6$ is not necessarily hermitian; we also show that it cannot be (strictly) nearly K\"{a}hler. 

In either case there is a crucial difference with the structures arising in supersymmetric heterotic compactifications: The Killing spinor equations (\ref{KS}), (\ref{KS0}) below are of the same form as those for the vanishing of the gravitino and dilatino variation in the heterotic theory, provided we set the background field $E^{\circ}_m$ to zero. For $\theta\neq0$,  
this would impose the additional condition that the manifold should be conformally balanced, which is indeed a necessary condition in supersymmetric heterotic compactifications. However as already mentioned for the rigid supersymmetric theories considered here the conformally balanced condition need not be imposed. Indeed as we will see in the following, a rigid (1,1)-supersymmetric theory can be defined on the round $\mathbb{S}^3\times\mathbb{S}^3$ which is hermitian, strong K\"{a}hler with torsion but not conformally balanced.
For $\theta=0$, the manifolds need not to be hermitian, while hermiticity is a necessary condition in supersymmetric heterotic compactifications.

As a special case of our formalism we obtain pure super Yang-Mills (SYM) 
on a Calabi-Yau threefold $\mathcal{M}_6$ by setting to zero all background fields except for the metric. Applying the 
localization procedure then yields the result that the SYM path integral 
localizes on stable holomorphic bundles on $\mathcal{M}_6$.

The rest of the paper is organized as follows: In section~\ref{sec:kse1},
we present the Killing spinor equations obtained from off-shell supergravity 
in the superconformal tensor calculus. In section~\ref{sec:vector}, we construct
the Lagrangian and supersymmetry transformation rules for Yang-Mills theory
on curved Riemannian manifolds $\mathcal{M}_6$ with background fields 
admitting solutions to the Killing spinor equations. 
Section~\ref{sec:tensor} extends the construction to include a number of 
tensor multiplets interacting by Chern-Simons couplings
with the vector fields.
In section~\ref{sec:kse2} we analyse the constraints to be satisfied by the 
background fields in order to admit non-trivial solutions. 
We reformulate the conditions for minimal supersymmetry as a set of necessary 
and sufficient conditions on a suitable $SU(3)$ structure. 
Moreover, we examine the conditions for (1,1) and (1,2) extended supersymmetry. The formalism is illustrated 
with a number of examples, including Calabi-Yau (CY) threefolds, 
the round $\mathbb{S}^3\times\mathbb{S}^3$ and
$T^2$ bundles over noncompact K3's. 
In section~\ref{sec:local} we apply the localization procedure to SYM theory
defined on a CY  ${\cal M}_6$ and show that 
the path integral localizes on stable holomorphic 
bundles over ${\cal M}_6$.


\section{Killing spinor equations I}
\label{sec:kse1}


Our starting point is the off-shell formulation of six-dimensional supergravity 
obtained from superconformal tensor calculus with the dilaton Weyl multiplet
coupled to a linear multiplet after particular gauge 
fixing~\cite{Bergshoeff:1985mz,Coomans:2011ih,Bergshoeff:2012ax}.
The resulting bosonic fields including the space-time metric
will constitute the background fields in the 
rigidly supersymmetric field theories to be constructed in this paper.
 As discussed in the introduction, when deriving rigid supersymmetric theories from
supergravity one has to impose the vanishing of 
the supersymmetry variation of all fermionic fields from the supergravity multiplet.
This defines the Killing spinor and poses constraints on the bosonic background fields.
In our conventions (cf.~appendix~\ref{app:6D})
and switching from Minkowsi to Euclidean signature,
the Killing spinor equations imposing the vanishing of the gravitino fields' variation, read
\bea
D_m \epsilon +\frac{1}{8} \, H^{\circ}_{mkl}\,\gamma^{kl}\,\epsilon
&=& 0
\;,
\nonumber\\
D_m \tilde\zeta -\frac18\, \tilde\zeta \gamma^{kl} H^{\circ}_{mkl}
&=& 0
\;,
\label{KS}
\eea
with covariant spinor derivatives 
\bea
D_m \epsilon &\equiv& \partial_m \epsilon 
+ \frac14\omega^{\circ}_m{}^{ab}\, \gamma_{ab} \,\epsilon +V^{\circ}_m\,  \epsilon
\;,
\label{covspin}
\eea
etc., a background vector field $V^{\circ}_m$, and a closed 
three-form $H^{\circ}_{mnk}=3\partial^{\vphantom{r}}_{[k}B^\circ_{mn]}$\,.
In passing from Minkowski to Euclidean signature, we have replaced the 
symplectic Majorana-Weyl spinors by complex Weyl spinors $\epsilon$, $\zeta$,
of positive chirality, 
thus reducing the manifest
R-symmetry from $Sp(1)$ to $U(1)$. In particular, unlike in Minkowski signature, 
the spinors $\epsilon$ and $\zeta$  are no longer related by complex conjugation.
If the Killing spinor equations (\ref{KS}) descend from a supergravity theory in which
the R-symmetry is gauged, there are additional Killing spinor equations to
be taken into account.
Namely, imposing the vanishing of the remaining fermion variations in that case furthermore implies
\bea
\gamma^m \epsilon\, \partial_m L^\circ - \frac16 L^\circ H^\circ_{mnk}\,\gamma^{mnk}\epsilon
 +\frac12  E_m^{\circ} \gamma^m \epsilon   &=& 0
\;,
\nonumber\\
\tilde\zeta \gamma^m \, \partial_m L^\circ + \frac16 L^\circ H^\circ_{mnk}\,\tilde\zeta \gamma^{mnk}
 -\frac12  E_m^{\circ} \tilde\zeta \gamma^m    &=& 0
\;,
\label{KS0}
\eea
where $L^\circ$ denotes the background dilaton and $E^{\circ}_{m}$
is the Hodge dual of the background five-form field strength\footnote{
Since we are in curved background geometry, the completely antisymmetric 
$\varepsilon^{mnklpq}$ is understood to be defined as a tensor, i.e.\
it carries an implicit factor of $\sqrt{g^\circ}^{-1}$.}
\bea
E^{\circ\,m} &=& \frac1{24}\,\varepsilon^{mn_1\dots n_5}\,\partial^{\vphantom{\circ}}_{n_1} E^{\circ}_{n_2\dots n_5}
\;.
\label{backE}
\eea
In the following we will construct rigidly supersymmetric field theories on backgrounds that allow for non-trivial solutions of the Killing spinor equations (\ref{KS}), (\ref{KS0}). 
In slight abuse of standard notation, we will refer to a background as $N=(p,q)$ supersymmetric when 
equations (\ref{KS}), (\ref{KS0}) admit
$p$ independent solutions $\epsilon$ and $q$ independent solutions~$\tilde\zeta$.\footnote{
This notation is not to be confused with the standard notion of chirality
in Minkowski space, since here $\epsilon$ and $\zeta$ are of the same chirality.}
Flat space with vanishing background fields would thus correspond to $N=(4,4)$.
Note the unlike the analogous structures in four dimensions~\cite{Cremmer:1982en,Festuccia:2011ws} 
the Killing spinor equations (\ref{KS}), (\ref{KS0}) for $\epsilon$ and $\tilde\zeta$ are entirely decoupled.
For the study of minimal supersymmetry we can thus consistently set $\tilde\zeta=0$.

While we will give a full-fledged analysis of the Killing spinor equations in 
section~\ref{sec:kse2} below, let us finish this section by deriving a few immediate consequences
for backgrounds that are at least $N=(1,1)$ supersymmetric.
Straightforward computation gives rise to the relations
\bea
\nabla_m \xi_n  &=&
-\frac12 H^\circ_{mnk} \,\xi^k
\;,\qquad
\nabla_m \xi_{p q r}  ~=~
\frac32 H^\circ_{m n [p} \xi^{\vphantom{m}}_{q r]}{}^{n}   
\;,
\label{dxi}
\eea
for the Killing spinor bilinears 
\bea
\xi^m~\equiv~-\tilde\zeta \gamma^m \epsilon\;,\qquad
\xi^{mnk}~\equiv~-\tilde\zeta \gamma^{mnk} \epsilon\;.
\label{xi}
\eea
In particular, the first equation shows that 
$\xi^m$ is a Killing vector of the background metric.
Furthermore, integrability of the Killing spinor equations (\ref{KS}) yields
\bea
0&=&
\frac18 R^\circ_{mn}{}^{ab}\gamma_{ab} \epsilon +\frac12 V^\circ_{mn}  \epsilon
+\frac{1}{8} \,\gamma^{kl}\,\epsilon\, \nabla^{\vphantom{m}}_{[m} H^{\circ}_{n]kl}
-\frac{1}{16} \, H^{\circ}_{mp}{}^k\,H^{\circ}_{nqk}\,\gamma^{pq}\, \epsilon
\;,\nonumber\\
0&=&
\frac18 R^\circ_{mn}{}^{ab}\,\tilde\zeta\gamma_{ab}  
+\frac12 V^\circ_{mn} \tilde\zeta 
+\frac{1}{8} \,\tilde\zeta\gamma^{kl}\, \nabla^{\vphantom{m}}_{[m} H^{\circ}_{n]kl}
-\frac{1}{16} \, H^{\circ}_{mp}{}^k\,H^{\circ}_{nqk}\,\tilde\zeta\gamma^{pq}
\;.
\label{integrab}
\eea
Contracting both equations with $\gamma^n$ and summing them up implies
\bea
\xi^{nkl}\, \nabla^{\vphantom{m}}_m H^\circ_{nkl}  &=& 
-\frac32 \, H^\circ_{mnp} H^\circ_{kl}{}^p\,\xi^{nkl} -12\, V^\circ_{mn}\,\xi^n
\;.
\eea
Together with (\ref{dxi}) we obtain
\bea
\partial_m\,\xi_H&=& -2 \, V^\circ_{mn}\,\xi^n
\;,
\qquad
{\rm for}\;\; \xi_H~\equiv~ \frac16\,\xi^{nkl} H^\circ_{nkl}
\;,
\label{const}
\eea
and anticipating the result of the full analysis that the background field strength
$V^\circ_{mn}$ vanishes (cf.~equation~(\ref{v}) below),
shows that the scalar combination $\xi_H$ is in fact constant.


\section{Vector multiplets}
\label{sec:vector}


We will now construct rigidly supersymmetric gauge theories on a background that 
admits non-trivial solutions of the Killing spinor equations (\ref{KS}), (\ref{KS0}).
The resulting couplings are motivated by the coupling~\cite{Bergshoeff:2012ax} 
of a single off-shell vector multiplet to the off-shell supergravity after integrating
out the auxiliary fields. 
However, since the spinor properties,
in particular their R-symmetry representation content 
change with the passage from Minkowski to Euclidean space, we will perform the full
construction from scratch.

Consider a set of vector multiplets $\{A_m^r, \lambda^r, \nu^r\}$ , labeled by an index $r$,
described by a Lagrangian
\bea
{\cal L}_{\rm vec} &=& 
-\frac14\,F_{mn}{}^r F^{mn\,r}-2 \tilde\nu^r \gamma^m D_m \lambda^r + \theta^r\,E^{\circ\,m} A^r_m
\nonumber\\
&&{}
+\frac{i}{16}\,\varepsilon^{mnklpq}\,B^{\circ}_{mn}\,F_{kl}{}^r\,F_{pq}{}^r
+\frac{1}{12} \tilde\nu^r \gamma^{mnk} H_{mnk}^{\circ}\,\lambda^r
\;.
\label{Lag_vec}
\eea
Here, ${F}_{mn}{}^r \equiv 2 \partial^{\vphantom{r}}_{[m} A_{n]}{}^r - f_{st}{}^r A_m{}^s A_n{}^t$
is the standard Yang-Mills field strength with structure constants $f_{st}{}^r$, 
and the spinor covariant derivatives are defined 
as in (\ref{covspin}).
 $B^\circ_{mn}$ is the background two-form potential, 
 the background field~$E^{\circ\,m}$ has been defined in (\ref{backE})
 and satisfies $\nabla_m E^{\circ\,m} =0$.
 The~$\theta^r$ denote a set of coupling constants which single out one of the vector fields
 and descend from an R-symmetry gauging of the underlying supergravity~\cite{Bergshoeff:2012ax}.
Accordingly, their gauge invariance requires an abelian factor within the Yang-Mills gauge group.
(For $\theta^r=0$ on the other hand there is no restriction on the gauge group.)
Again, the spinors $\lambda^r$ and $\nu^r$ are not related by complex conjugation,
such that the variation of the vector fields and the Lagrangian are in fact complex,
as usual in Euclidean supersymmetry, cf.~\cite{Osterwalder:1973dx,Nicolai:1978vc,vanNieuwenhuizen:1996tv}, 
see also the discussion in section~\ref{sec:local}.

It is straightforward to check that the Lagrangian (\ref{Lag_vec}) is invariant under
the supersymmetry transformations rules
\bea
\delta A_m{}^r &=& \tilde\nu^r \gamma_m \epsilon - \tilde\zeta \gamma_m \lambda^r
\;,\nonumber\\
\delta \lambda^r &=& \frac14\,\gamma^{mn}\epsilon\,F_{mn}{}^r
-\theta^r L^\circ\,\epsilon
\;,\nonumber\\
\delta \tilde\nu^r &=& -\frac14\,\tilde\zeta\gamma^{mn}\,F_{mn}{}^r
+\theta^r L^\circ\,\tilde\zeta
\;,
\label{susy_vector}
\eea
provided the
supersymmetry parameters satisfy the Killing spinor equations (\ref{KS}), (\ref{KS0}).
The latter equations are only required in case the $\theta^r$ are non-vanishing.
For $\theta^r=0$, the only condition on the supersymmetry parameters are equations~(\ref{KS}).
Let us note that variation of the gauge connection in the Dirac term a priori
gives rise to a quartic fermion term in the variation of (\ref{Lag_vec}) 
\bea
\delta_4{\cal L} &\propto&
f_{rst}\, (\tilde\nu^r \gamma^m \lambda^t) \,(\tilde\nu^s \gamma_m \epsilon)
\;.
\label{quartic}
\eea
Fierzing in $\lambda^t) \,(\tilde\nu^s$ shows that
\bea
f_{rst}\, (\tilde\nu^r \gamma^m \lambda^t) \,(\tilde\nu^s \gamma_m \epsilon)
&=&
 f_{rst}\, (\tilde\nu^s \gamma^k \lambda^t) (\tilde\nu^r  \gamma_k \epsilon)
\;,
\eea
upon using that $\gamma^m \gamma_{kln} \gamma_m=0$.
By virtue of antisymmetry of the structure constants, also the quartic variation (\ref{quartic}) thus vanishes.

It is instructive to work out the $N=(1,1)$ algebra of supersymmetry transformations (\ref{susy_vector})
which takes the form\footnote{
Part of these and the following
calculations have been facilitated by use of the computer algebra system Cadabra~\cite{Peeters:2006kp,Peeters:2007wn}.} 
\bea
{}[\delta_{\tilde\zeta},\delta_\epsilon] \,A_m{}^r &=& 
{\cal L}_\xi A_m^r + D_m (-\xi^n A_n^r)
\;,
\nonumber\\
{}[\delta_{\tilde\zeta},\delta_\epsilon] \, \lambda^r &=& 
{\cal L}_\xi \lambda^r 
+\frac1{2}\, \xi_H \, \lambda^r
\;,
\nonumber\\
{}[\delta_{\tilde\zeta},\delta_\epsilon] \, \tilde\nu^r &=& 
{\cal L}_\xi \tilde\nu^r 
-\frac1{2}\, \xi_H \, \tilde\nu^r
\;,
\label{susy_algebra}
\eea
with parameters $\xi^n$ and $\xi_H$ defined in (\ref{xi}) and (\ref{const}),
respectively,
and defining the standard Lie derivative on spinor fields
\bea
{\cal L}_\xi \chi^I &\equiv&
 \xi^m \nabla_m \chi^I + \frac14\left(\nabla_m \xi_n\right) \gamma^{mn}\,\chi^I
 ~=~
\xi^m \nabla_m \chi^I - \frac18\,\xi^k H^\circ_{kmn}\gamma^{mn}\,\chi^I
\;.
\eea
Interestingly, we find that the algebra closes not only into the standard
translations and gauge transformation but also into a global $U(1)$ acting
on the fermions with  parameter $\xi_H$ that is constant due to (\ref{const}).\footnote{As we will see in the following the full analysis of the Killing spinor equations shows that the constant $\xi_H$ vanishes in the case of backgrounds for which $E_m^{\circ}=0$, but is generically non-vanishing. However, as the Lagrangian (\ref{Lag_vec})
possesses the global $U(1)$ appearing under closure on the r.h.s.\ of (\ref{susy_algebra}),
the form of the algebra in any case is consistent, whether or not  $\xi_H$ is zero.}

To summarize, the Lagrangian (\ref{Lag_vec}) defines non-abelian 
gauge theory on a curved Euclidean background with supersymmetry parameters
defined as solutions of the Killing spinor equations (\ref{KS}), (\ref{KS0}).


\section{Tensor multiplets}
\label{sec:tensor}


We will now generalize the construction of the previous section to also include
tensor multiplets.
To this end, we will first derive the field equations for a set of free tensor multiplets
on curved background and then proceed to introduce interactions with the
vector multiplets. Since supersymmetry is expected to impose self-duality
equations for the tensor fields that do not allow for a standard Lagrangian description,
we will perform the entire construction on the level of the field equations.
Moreover, since there is no off-shell formulation for self-dual tensor fields 
in six dimensions, their couplings on a curved background cannot directly be read off
from an underlying supergravity but have to be constructed from scratch.

Let us consider $n$ tensor multiplets $\{\phi^I, B_{mn}^I, \chi^I, \tilde\psi^I\}$, 
labelled by an index $I$, with supersymmetry transformations
\bea
\delta \phi^I &=& \tilde\psi^I \epsilon + \tilde\zeta \chi^I
\;,\nonumber\\
\delta B_{mn}{}^I &=& \tilde\psi^I\,\gamma_{mn}\,\epsilon
- \tilde\zeta \gamma_{mn} \chi^I
\;,\nonumber\\
\delta \chi^I &=& \frac12 \gamma^m \epsilon\, \partial_m \phi^I
+\frac1{24} \,\gamma^{mnk} \epsilon\left(
H_{mnk}{}^I
-H^{\circ}_{mnk}\,\phi^I\right)
\;,\nonumber\\
\delta \tilde\psi^I &=& -\frac12 \tilde\zeta \gamma^m\, \partial_m \phi^I
+\frac1{24} \,\tilde\zeta\gamma^{mnk} \left(
H_{mnk}{}^I
-H^{\circ}_{mnk}\,\phi^I\right)
\;. 
\label{susy_tensor}
\eea
where
\bea
H_{mnk}{}^I &\equiv& 3 \partial_{[m} B_{nk]}{}^I\;,
\eea
denotes the abelian tensor field strength which decomposes into its selfdual and anti-selfdual part 
according to
\bea
H_{mnk}^{\pm} &\equiv&
\frac12\Big(
H_{mnk}
\pm \frac{i}{3!}\,\varepsilon_{pqrmnk}\,H^{pqr}
\Big)
\;.
\label{Hplus}
\eea

Closure of the supersymmetry algebra on the tensor fields gives rise to
the self-duality equation
\bea
(H_{mnk}{}^I)^{(+)} - \phi^I ({H}^{\circ}_{mnk}{})^{(+)} &=& 0
\;,
\label{Hselfdual}
\eea
relating the tensor field strength to the background three-form.
Under the supersymmetry transformations (\ref{susy_tensor}), 
the self-duality equation transforms into the fermionic field equations
\bea
0&=&
D_m\tilde\psi^I \gamma^m + \frac{1}{24}\,\tilde\psi^I \gamma^{mnk}\,H_{mnk}^{\circ}
\;,
\nonumber\\
0&=&
\gamma^m D_m\chi^I - \frac{1}{24}\,H_{mnk}^{\circ}\,\gamma^{mnk}\,\chi^I
\;,
\label{eom_psi}
\eea
provided the supersymmetry parameters $\epsilon$, $\tilde\zeta$ satisfy the 
Killing spinor equations (\ref{KS}).
In turn, these equations under supersymmetry transform
into the scalar field equation
\bea
0&=&
\nabla^\mu\partial_\mu\phi^I +\frac1{6}\,(H^\circ \cdot H^I{}^{(-)})  
-\frac1{12}\,(H^\circ \cdot H^\circ){}\,\phi^I  
\;,
\eea
where again (\ref{KS}) is required.
For consistency, we may check the $N=(1,1)$ algebra of supersymmetry transformations~(\ref{susy_tensor})
which close into
\bea
{}[\delta_{\tilde\zeta},\delta_\epsilon] \,\phi^I &=& {\cal L}_\xi \phi^I
\;,
\nonumber\\
{}[\delta_{\tilde\zeta},\delta_\epsilon] \,B_{mn}{}^I &=& {\cal L}_\xi B_{mn}{}^I 
+2\, \partial_{[m} \Lambda_{n]}{}^I\;,
\nonumber\\
{}[\delta_{\tilde\zeta},\delta_\epsilon] \, \chi^I &=& 
{\cal L}_\xi \chi^I
+\frac1{2}\, \xi_H \, \chi^I
\;,
\label{susy_algebra_ten}
\eea
with the same translation and $U(1)$ parameters $\xi^m$, $\xi_H$,
as in (\ref{susy_algebra}), and the additional tensor gauge parameter $\Lambda_m{}^I \equiv
-\xi^n B_{nm}{}^I+ \xi_m \phi^I$.

Finally, we may couple the system of tensor multiplets to the non-abelian vector fields
introduced in the previous section by introducing Chern-Simons interactions 
of the form
\bea
H_{mnk}{}^I &\rightarrow& {\cal H}_{mnk}{}^I
  ~\equiv~ 3 \partial^{\vphantom{m}}_{[m} B_{nk]}{}^I +
6 \, d^I_{rs}  A_{[m}{}^r \partial^{\vphantom{r}}_n A_{k]}{}^s
- 2 f_{pq}{}^s d^I_{rs} A_{[m}{}^r A_n{}^p A_{k]}{}^q
\;,
\label{defF}
\eea
parametrized by a constant gauge invariant tensor $d^I_{rs}=d^I_{(rs)}$.
For simple gauge groups, this implies that $d^I_{rs}=d^I\delta_{rs}$ is expressed in terms of the 
Cartan-Killing form $\delta_{rs}$.
The modified tensor field strength ${\cal H}_{mnk}{}^I$ satisfies
\bea
\partial_{[m} {\cal H}_{nkl]}{}^I &=&
\ft32\, d^I_{rs} \, {\cal F}_{[mn}{}^r {\cal F}_{kl]}{}^s
\;,
\nonumber\\[.5ex]
\delta\, {\cal H}_{mnk}{}^I &=& 3\, \partial_{[m} 
\left(\delta B_{nk]}{}^I-2 A_{n}{}^r \,\delta A_{k]}{}^s\,d^I_{rs}\right)
+6\, d^I_{rs} \, {\cal F}_{[mn}{}^r \,\delta A_{k]}{}^s
\;.
\eea
In presence of $d^I_{rs}$ the supersymmetry transformation rules (\ref{susy_tensor}) 
change to
\bea
\delta B_{mn}{}^I &=& \tilde\psi^I\,\gamma_{mn}\,\epsilon
- \tilde\zeta \gamma_{mn} \chi^I
+2 d^I_{rs}\,A_{[m}{}^r \,\delta A_{n]}{}^s
\;,\nonumber\\[.5ex]
\delta \chi^I &=& \frac12 \gamma^m \epsilon\, \partial_m \phi^I
+\frac1{24} \,\gamma^{mnk} \epsilon\left(
H_{mnk}{}^I
-H^{\circ}_{mnk}\,\phi^I\right)
\nonumber\\
&&{}
+\frac12 d^I_{rs}\gamma^\mu\lambda
\left(\tilde\nu\gamma_\mu\epsilon
-\tilde\zeta\gamma_\mu\lambda\right)
\;,
\label{susy_CS}
\eea
and similar for $\tilde\psi$\,, which again can be verified by
checking closure of the algebra.
The field equations change into
\bea
0 &=& (H_{mnk}{}^I)^{(+)} - \phi^I ({H}^{\circ}_{mnk}{})^{(+)} 
+d^I_{rs} \tilde\nu^r \gamma_{mnk} \lambda^s 
\;,\nonumber\\
0&=&
\gamma^m D_m\chi^I - \frac{1}{24}\,H_{mnk}^{\circ}\,\gamma^{mnk}\,\chi^I
-\frac12\,d_{rs}^I F^r_{mn} \gamma^{mn} \lambda^s
-2 \,d_{rs}^I \,\theta^r L^\circ \lambda^s
\;,
\nonumber\\
0&=&
\nabla^\mu\partial_\mu\phi^I +\frac1{6}\,(H^\circ \cdot H^I{}^{(-)})  
-\frac1{12}\,(H^\circ \cdot H^\circ){}\,\phi^I \nonumber\\
&&{} +4\,d_{rs}^I \,\Big(
\,\frac18\,F_{mn}{}^rF^{mn}{\,}^s
+\tilde\nu^r \gamma^{m}\partial_m  \lambda^s
+\theta^r\theta^s (L^\circ)^2 \Big)
\;,
\label{eom_tensors}
\eea
and can be checked to vary into each other under supersymmetry,
provided the supersymmetry parameters $\epsilon$, $\tilde\zeta$ satisfy the 
Killing spinor equations (\ref{KS}), (\ref{KS0}).
This generalizes the field equations found in flat space~\cite{Bergshoeff:1996qm}
for a tensor/vector system interacting by Chern-Simons couplings to curved background. 
In deriving this result, we had to use the gaugino field equation coming from (\ref{Lag_vec}).
I.e.\ unlike in flat space where the YM sector can be kept off-shell
in the presence of Chern-Simons couplings~\cite{Bergshoeff:1996qm}, 
the non-vanishing background field $H^\circ_{mnk}$
relates the dynamics of the vector and tensor sector.

It would be interesting to further generalize the interactions by allowing for
minimal couplings between vector and tensor fields with the latter charged under 
a non-abelian gauge group. The corresponding supersymmetric system in flat space
has been constructed in~\cite{Samtleben:2011fj,Samtleben:2012mi} and allows for an action modulo the
standard subtleties concerning self-dual three-forms. Replacing the constant supersymmetry
parameters of that system by solutions to the Killing spinor equations (\ref{KS})
should lead to a deformed version of these tensor/vector interactions on curved space.
It would also be interesting to analyze if the curved space models can be generalized
to include couplings to hypermultiplets, by applying the same procedure to
the flat space models of~\cite{Sierra:1983uh,SSWhyp}.


\section{Killing spinor equations II}
\label{sec:kse2}


So far, we have derived the couplings and field equations for vector and tensor multiplets on a curved background
under the assumption that the supersymmetry parameters satisfy the Killing spinor equations (\ref{KS}), (\ref{KS0}).
In this section, we will analyze these Killing spinor equations and derive the resulting constraints for the background fields,
in particular the background geometry.
We obtain the necessary and sufficient conditions for the background geometry in order
to admit at least one Killing spinor, i.e.\ for the theory to have (1,0) supersymmetry.
In subsections~\ref{sec:11}, \ref{sec:12} we derive stronger conditions for the
existence of several supercharges.
As this section deals exclusively with the background structure we will for notational simplification omit the superscript `$^\circ$' which so far has
distinguished the background structures from the matter fields. I.e.\ (for this section only)
we pass to $H^\circ \rightarrow H$, $L^\circ \rightarrow L$, etc.

\subsection{(1,0) supersymmetry}\label{sec:10}

The Killing spinor equation (\ref{KS}) can be thought of as a parallel-transport equation, $\nabla'\epsilon=0$, with respect to a metric-compatible connection $\nabla'$ with torsion given by the background three-form $H$. Therefore the holonomy of $\nabla'$ (in the spinor representation) is a subgroup of $G$, the stabiliser group of $\epsilon$, and $\mathcal{M}_6$ admits a $G$-structure \cite{joyce}. Since $\epsilon$ transforms in the ${\bf 4}$ of the structure group $Spin(6)\cong SU(4)$ of the Riemannian spin manifold $\mathcal{M}_6$, the stabiliser is $G=SU(3)$ as can be seen by the decomposition ${\bf 4}\rightarrow {\bf 3}\oplus{\bf 1}$ under $SU(4)\rightarrow SU(3)$. Hence $\mathcal{M}_6$ must admit an $SU(3)$ structure.

The topological obstruction for an oriented Riemannian six-dimensional manifold $\mathcal{M}_6$ to admit an $SU(3)$ structure is that it should be spin (see e.g. \cite{Witt:2006tq}). Since in the present paper we are assuming that $\mathcal{M}_6$ is spin, there is no additional topological condition imposed by the existence of an $SU(3)$ structure on $\mathcal{M}_6$. However, as we will see in the following, supersymmetry imposes additional geometrical conditions in the form of constraints on the torsion classes of the $SU(3)$ structure. In order to systematically analyze these constraints we will now reformulate the  $SU(3)$ structure in terms of certain globally defined forms on $\mathcal{M}_6$.

From equation (\ref{KS})  it can easily be seen  that either $\epsilon$ vanishes identically, or it is nowhere-vanishing. 
Assuming that $\epsilon$ is not identically zero, let us parametrize:
\be\label{eah}
\epsilon=e^{A}\eta~,
\ee
where $\eta$ is a unimodular Weyl spinor of positive chirality,
\be{}
\eta^{\dagger}\eta=1~,
\ee
and $A$ is a function on the six-dimensional manifold. For the purposes of this 
section we will assume that $\eta$ is a commuting spinor: we have the freedom to make this choice since the Killing spinor equations are linear in $\epsilon$.

The existence of the unimodular spinor $\eta$ allows us to define the bilinears
\eq{\spl{\label{o}
iJ_{mn}&:=\eta^{\dagger}\gamma_{mn}\eta \\
\Omega_{mnp}&:=\tilde{\eta}\gamma_{mnp}\eta~,
}}
where our spinor conventions are explained in detail in appendix \ref{app:6D}. In particular, it follows from the properties of Weyl spinors in six Euclidian dimensions that  $J$ is a real two-form and 
$\Omega$ is a complex three-form which is imaginary self-dual.

Using the Fierz identities it can further be shown 
that $J_m{}^n$ (where we have raised one index with the metric) defines an almost complex structure:
\be\label{osa}
J_m{}^nJ_n{}^p=-\delta_m{}^p~,
\end{equation}
with respect to which $\Omega$ is $(3,0)$:
\be
(\Pi^+)_m{}^n\Omega_{npq}=\Omega_{mpq}; ~~~~~~ 
(\Pi^-)_m{}^n\Omega_{npq}=0  ~,
\end{equation}
where 
\be
(\Pi^\pm)_m{}^n:=\frac{1}{2}(\delta_m{}^n\mp i J_m{}^n)
\la{projectors}~,
\end{equation}
are the projection operators onto the (1,0) and (0,1) parts. 
Moreover, it can be seen that
\beal
\Omega\wedge J&=0\nn\\
\Omega\wedge\Omega^*&=\frac{4i}{3}J^3~.
\end{align}
The globally 
defined forms $J$, $\Omega$ subject to the above conditions can be seen to 
specify an $SU(3)$ structure on $\mathcal{M}_6$ \cite{chiossi}.

The intrinsic torsion parametrizes the 
failure of the spinor $\eta$ to be covariantly constant. 
In the case of an $SU(3)$-structure manifold $\mathcal{M}_6$, 
the intrinsic torsion  decomposes into five modules (torsion classes) 
${W}_1, \dots, {W}_5$: 
\beal
\nabla_m\eta&=\frac{1}{2}\left(W_{4m}^{(1,0)}+W_{5m}-\mathrm{c.c} \right)\eta\nn\\
&+\frac{1}{16}\left(
4W_1g_{mn}-2W_4^p\Omega_{pmn}+4iW_{2mn}-iW_{3mpq}\Omega^{pq}{}_n
\right)\gamma^n\eta^c ~,
\label{torsionclasses}
\end{align}
where $W_1$ is a complex scalar, $W_2$ is a complex (1,1)-traceless form, $W_3$ is a real traceless $(2,1)+(1,2)$ form, $W_4$ is a real one-form and $W_5$ is a (1,0) form. 
In terms of $SU(3)$ representations,
\be
W_1\sim {\bf 1}\oplus  {\bf {1}};~~~~~ W_2\sim {\bf 8}\oplus  {\bf {8}}; 
~~~~~ W_3\sim {\bf 6}\oplus  {\bf \bar{6}}; 
~~~~~ W_4\sim {\bf 3}\oplus  {\bf \bar{3}};  ~~~~~ W_5\sim {\bf {3}}~.
\end{equation}

Equivalently these torsion classes appear  
in the $SU(3)$ decomposition of the exterior derivatives of $J$, $\Omega$. Explicitly we have,
\beal
dJ&=\frac{3}{2}\mathrm{Im}(W_1^*\Omega)+W_4\wedge J+W_3\;,\nn\\
d\Omega&= W_1 J\wedge J+W_2 \wedge J+\Omega\wedge W_5^* ~.
\label{torsionclassesb}
\end{align}
Intuitively, the intrinsic torsion characterizes the geometry of 
$\mathcal{M}_6$ and parameterizes the 
failure of the manifold to be of special holonomy -- which can also 
be thought of as the failure of the closure of $J$, $\Omega$.

As a final ingredient before we proceed to the analysis of the Killing spinor equations, we will need the decomposition of the background three-form $H$ from (\ref{KS}) with respect to the reduced structure group $SU(3)$. 
Explicitly we have
\be
H_{mnp}=\frac{1}{48}\Omega_{mnp}H^{(0)}
+(\tilde{H}_{mnp}^{(2,1)}+\frac{3}{4}H^{(1,0)}_{[m}J_{np]}  ) 
+{\rm c.c.}~,
\la{hexp}
\end{equation}
where the normalization above has been chosen so that
\beal
H^{(0)}&=\Omega^{*mnp}H_{mnp}\;,\nn\\
H^{(1,0)}_m&=(\Pi^+)_m{}^sH_{snp}J^{np}~.
\end{align}
In terms of $SU(3)$ representations we have,
\be
H^{(0)}\sim {\bf 1}; 
~~~~~ H^{(1,0)}\sim {\bf 3}; ~~~~~ 
H^{(0,1)}\sim {\bf \bar{3}}; ~~~~~ 
\tilde{H}^{(2,1)}\sim {\bf {6}}; ~~~~~ \tilde{H}^{(1,2)}\sim {\bf \bar{6}}~.
\end{equation}

Plugging (\ref{torsionclasses}), (\ref{hexp}) into the Killing spinor equation (\ref{KS}) using (\ref{a1}) can be seen to impose the following constraints on the torsion classes:
\beal\label{tca}
W_2&=0\;,\nn\\
2W_4^{(1,0)}+W_5&=0~.
\end{align}
Moreover it follows that the background vector field is exact,
\be\label{v}
dA+V=0~,
\ee
and  the irreducible components of $H$ are determined in terms of the 
torsion classes. Explicitly, in form notation:
\beal\label{tcb}
H^{(0)}&=-12W_1^*\;,\nn\\
H^{(1,0)}&=-4iW_4^{(1,0)}\;,\nn\\
\tilde{H}^{(1,2)}&=iW_3^{(1,2)}
~.
\end{align}

\subsubsection{$\theta^r\not=0$}
\label{sec:Lnot0}

For the remaining analysis, it will make a crucial difference whether 
the parameters $\theta^r$ in the Lagrangian (\ref{Lag_vec}) are vanishing or not. 
Only in the latter case do we have to impose the Killing spinor equations (\ref{KS0}) on
the supersymmetry parameters. This leads to further restrictions on the background geometry,
to be discussed in this section. The case $\theta^r=0$ is examined separately in section \ref{sec:L0}.

Without loss of generality, we may assume in this section
the background field $L$ to be non-vanishing.
Otherwise, it is straightforward to see that
the Killing spinor equation (\ref{KS0}) implies that also $E_m=0$ (taking into 
account that $E_m$ is real). For the Lagrangian (\ref{Lag_vec}) and the supersymmetry
transformation rules (\ref{susy_vector}), this amounts to setting $\theta^r=0$, and brings us back to the situation
discussed in the next section.
With the field $L$ non-zero, the second Killing spinor equation (\ref{KS0}) 
implies  that the scalar component of $H$ must vanish:
\be{}\label{hcond0}
H^{(0)}=0~.
\ee
As a consequence of the above and (\ref{tcb}) there is 
now an additional constraint on the torsion classes:
\beal\label{tcc}
W_1&=0~.
\end{align}
Taking the above into account, it follows that  the $H$ field is given in terms of  the torsions classes:
\eq{\label{tcbb}
H=-i\left[
(W_3^{(2,1)}-W_3^{(1,2)})+J\wedge(W_4^{(1,0)}-W_4^{(0,1)})
\right]
~.
}
By introducing explicit holomorphic coordinates, eq.~(\ref{tcbb}) can be rewritten as
\eq{\label{f1}
H=-i(\partial J-\bar\partial J)~,}
where $d=\partial+\bar\partial$ and we have noted that 
\eq{
\left(d J\right)^{(2,1)}=
\partial J~;~~~ \left(d J\right)^{(1,2)}=\bar\partial J
~.
}
It follows that the closure of $H$ can be written equivalently as:
\eq{
\partial\bar{\partial} J=0
~.}
Eq.~(\ref{tcbb}) can also be written as:
\eq{\label{hstar}
\star H=W_3-W_4\wedge J
~,}
where we have used 
\eq{
\varepsilon_{m_1\dots m_6}=-15J_{[m_1m_2}J_{m_3m_4}J_{m_5m_6]}
~.}
If $W_4$ is exact (which need not be true in general), $W_4=d\phi$, 
the manifold is called conformally balanced. 
In that case (\ref{hstar}) can be written as:
\eq{\label{f2}
\star H=e^{2\phi}d\left(e^{-2\phi} J\right)
~.}
Finally the background field $E_m$ is given by:
\be{}\label{econd}
E_m=-4LW_{4m}-2\partial_mL~,
\ee
where in addition  $\nabla^mE_m=0$ 
has to be imposed, cf. eq.~(\ref{backE}).

To summarize: the combined Killing spinor equations 
(\ref{KS}), (\ref{KS0}), the closure of the background three-form and the co-closure of the background one-form 
imply the following geometric constraints 
on $\mathcal{M}_6$:
\boxedeq{\spl{\label{tc}
W_1=W_2&=0\;,\\
W_4+\mathrm{Re}W_5&=0\;,\\
\partial\bar{\partial}J&=0\\
\nabla^m\left(\partial_mL+2LW_{4m}\right)&=0
~.
}}
Given an $SU(3)$-structure manifold obeying (\ref{tc}) there is no obstruction to specifying the profiles for the background fields $V$, $H$, $E$ so that the  remaining conditions (\ref{v}), (\ref{tcbb}), (\ref{econd}) are satisfied. In this sense  conditions (\ref{tc}) are {\it necessary and sufficient}   
for  the theory to possess rigid supersymmetry.

Conditions (\ref{tc}) have the following geometric interpretation:  
The vanishing of $W_1$, $W_2$ is equivalent to the condition that the almost complex structure is integrable, i.e. the condition that $\mathcal{M}_6$ is hermitian.\footnote{A manifold equipped with a Riemannian metric $g_{mn}$ and an almost complex structure $J_m{}^n$ is called {almost hermitian} if the almost complex structure is an orthogonal transformation with respect to the metric:
\eq{\label{osb}
J_m{}^pJ_n{}^qg_{pq}=g_{mn}~.
}
If moreover the almost complex structure  is integrable, the manifold is called {hermitian}.}
The second line of (\ref{tc}) can be rephrased as the statement of proportionality between  
the Lee forms $L_{\Omega}$ and $L_J$ of $\Omega$ and $J$ respectively. More explicitly the Lee forms are defined by:
\eq{\spl{
L_{J}&=J\lrcorner d J\\
L_{\Omega}&=\mathrm{Re}\left(\Omega\lrcorner d \Omega^*\right)
~,}}
where $\varphi\lrcorner\omega$ denotes the contraction of the $p$-form $\varphi$ into the $(p+q)$-form $\omega$,
\eq{
\varphi\lrcorner\omega
=\frac{1}{p!q!}\varphi^{m_1\dots m_p}\omega_{m_1\dots m_pn_1\dots n_q}dx^{n_1}\wedge\dots\wedge dx^{n_q}
~.}
On the other hand, taking (\ref{torsionclassesb}) into account, it follows that
\eq{
W_4=\frac{1}{2}J\lrcorner d J~;~~~
\mathrm{Re}W_5=\frac{1}{8}\mathrm{Re}\left(\Omega\lrcorner d \Omega^*\right)
~,}
hence the second line of (\ref{tc})   can be rewritten in terms of the Lee forms as:
\eq{
L_{J}+\frac{1}{4}L_{\Omega}=0
~.}
The third line of (\ref{tc}) is the condition that the Hermitian manifold $\mathcal{M}_6$ is strong 
K\"{a}hler with torsion (SKT). One immediate consequence of the SKT condition is that $\mathcal{M}_6$ cannot be special hermitian \cite{skt}.\footnote{A special hermitian six-manifold is a manifold which is both half-flat (and therefore can be lifted to a seven-dimensional manifold of $G_2$ holonomy \cite{hitc}) and hermitian. Equivalently,  a special hermitian six-manifold is a manifold which admits an  $SU(3)$ structure whose only non-zero torsion class is $W_3$.}

The first two lines of (\ref{tc}) also appear as necessary conditions for supersymmetric heterotic compactifications \cite{Strominger:1986uh,LopesCardoso:2002hd,Gauntlett:2003cy}. This is not surprising given the fact that the Killing spinor equations (\ref{KS}), (\ref{KS0}) are of the same form as those for the vanishing of the gravitino and dilatino variation in the heterotic theory, provided we set the background fields $V$, $E$ to zero and we identify $L=e^{-2\phi}$, where $\phi$ is the dilaton. Crucially, however, this would impose the additional condition that the torsion class $W_4$ is exact: $W_4=d\phi$ as follows from eq.~(\ref{econd}); in particular the manifold is constrained to be conformally balanced. 
We stress again that for the rigid supersymmetric theories considered here the conformally balanced condition need not be imposed.

\subsubsection{$\theta^r=0$}
\label{sec:L0}

With all parameters $\theta^r$ vanishing, the background fields $L$ and $E_m$ 
completely decouple form the Lagrangian (\ref{Lag_vec}) and
supersymmmetry transformation rules (\ref{susy_vector})
and equations (\ref{KS0}) do no longer have to be imposed on the supersymmetry parameters.
As a result, the constraints on the background geometry are relaxed w.r.t.\ the last section
and in particular the torsion class $W_1$ is no longer zero, cf.~(\ref{tcc}).
Equations (\ref{tcbb}), (\ref{hstar}) for the $H$ field now get modified to:
\eq{\label{tcbb0}
H=-\frac12 \mathrm{Re}\left(
W_1\Omega^*
\right)-i\left[
(W_3^{(2,1)}-W_3^{(1,2)})+J\wedge(W_4^{(1,0)}-W_4^{(0,1)})
\right]
~,
}
and:
\eq{\label{hstar0}
\star H=
\frac12 \mathrm{Im}\left(
W_1\Omega^*
\right)
+W_3-W_4\wedge J
~.}
Finally, the necessary and sufficient conditions can be summarized as
\boxedeq{\spl{\label{tc0}
W_2&=0\;,\\
W_4+\mathrm{Re}W_5&=0\;,\\
d\left[\mathrm{Re}\left(
W_1\Omega^*
\right)+2\star\left(
W_3-W_4\wedge J
\right)\right]&=0
~.}}
The fact that now $W_1$ is not necessarily zero implies that the almost complex structure is not integrable in general and  the manifold $\mathcal{M}_6$ is not necessarily hermitian. This is a major difference from the $L\neq0$ case discussed above. 
The second line above has the same interpretation as in section \ref{sec:Lnot0} as the proportianality between the 
Lee forms of $J$ and $\Omega$. The last line in (\ref{tc0}) is the condition of closure of $H$; for $W_1=0$ it reduces to the third 
line in (\ref{tc}).

As we will now show, the system (\ref{tc0}) does not admit as solution any (strictly) nearly K\"{a}hler manifold. Nearly K\"{a}hler 
manifolds are $SU(3)$-structure manifolds whose only nonvanishing torsion class is $W_1$. Equivalently, they can be defined as six-dimensional manifolds admitting a Killing spinor, $\nabla_m\eta=\frac{1}{4}W_1\gamma_m\eta^c$, cf.\ eq.~(\ref{torsionclasses}), so that they are Einstein, $R_{mn}=\frac{5}{4}|W_1|^2g_{mn}$, and the metric cone over $\mathcal{M}_6$ is a $G_2$-holonomy manifold \cite{baer}. In the case of a nearly K\"{a}hler  $\mathcal{M}_6$  it can immediately be seen from  eqs.~(\ref{torsionclassesb}) that $W_1$ is constant. Performing  a constant phase redefinition of $\Omega$ we can take $W_1=iw$, with $w$ a real constant. Eqs.~(\ref{torsionclassesb}) then reduce to:
\eq{
dJ=-\frac{3}{2}w\mathrm{Re}\Omega~,~~~d\mathrm{Im}\Omega=w J\wedge J
~,}
while $d\mathrm{Re}\Omega=0$ follows from the above. Moreover the background field $H$ is given by
\eq{
H=-\frac{1}{2}w\mathrm{Im}\Omega
~,
}
and it is not closed, unless $w=0$ and  $\mathcal{M}_6$ is a Calabi-Yau.

\subsection{(1,1) supersymmetry}\label{sec:11}

In the case of vanishing $\theta^r$, we have seen that the Killing spinor equations~(\ref{KS0})
do no longer have to be imposed. 
The remaining equations (\ref{KS}) are related by simple transposition, i.e.\ any solution $\epsilon$ of the first
equation in (\ref{KS}) defines a solution to the second equation by setting $\zeta\equiv\epsilon$.
In our notation, this means that the resulting theory in fact is $N=(1,1)$ supersymmetric rather than just $N=(1,0)$.\footnote{
Note that even 
when $\epsilon$, $\zeta$ are not linearly independent, they 
still have different action on the fields, cf.~(\ref{susy_vector}). 
The supersymmetry thus 
is (1,1) and not (1,0) or (0,1), which would correspond to one of the two spinor parameters being zero.} 

In contrast, for non-vanishing $\theta^r$, imposing (1,1) supersymmetry in general imposes 
further constraints on the background geometry, to be discussed in the rest of this section.
In this case, we have to look for 
necessary and sufficient conditions so that the Killing spinor eqs.~(\ref{KS}), (\ref{KS0}) 
are obeyed for a pair of not identically vanishing spinors~$\epsilon$, $\zeta$.
We have already seen that (1,0) supersymmetry imposes that the background 
gauge field is exact, cf.~eq.~(\ref{v}). 
Hence we may gauge away $V$ by a suitable redefinition of the dynamical fermions of the theory. 
Setting $V=0$ in (\ref{v}) then implies that $A$ is constant.  In this section we will take $A=0$ without loss of generality so that, cf.\ eq.~(\ref{eah}),
\eq{\epsilon=\eta~.} 
One immediate consequence of the Killing spinor equations (\ref{KS0}) is that by taking 
$\zeta=c\eta$, where $c$ is a constant,  the theory is (1,1) supersymmetric if and only if 
\eq{\label{conde}
E_m=0 
~,}
\addtocounter{footnote}{-1}
in addition to the conditions derived in section \ref{sec:Lnot0}.\footnotemark~
Combined with (\ref{econd}), eq.~(\ref{conde}) implies the exactness of the $W_4$ torsion class
\eq{W_4=d\phi~,}
where we have set $L=e^{-2\phi}$. 
Hence the manifold is constrained to be conformally balanced. As 
a consequence  it also follows that the 
scalar $\xi_H$ defined in (\ref{const}) vanishes.

More generally, any (commuting) spinor $\zeta$ of positive 
chirality can be written as:
\eq{\label{z}
\zeta=c\eta+\gamma^mK_m\eta^c
~,}
where $K$ is a (1,0)-form.\footnote{With a slight abuse of notation, we will use the same symbol for the form and the vector obtained from it by 
raising the index with the metric.} It can then be seen taking the formul\ae{} of appendix \ref{app:b} into account that the 
Killing spinor equations are equivalent to 
the following set of conditions:
\eq{\spl{\label{11cond}
c=0~~~\mathrm{or}~~~E_m&=0\\
K\lrcorner dL&=0\\
\nabla_{(m}K_{n)}&=0\\
dK+K\lrcorner H&=0\\
\frac{1}{4}\left(E^{(1,0)}-2\partial L\right)\wedge K-iL K\lrcorner W_3^{(2,1)}
&=0~,}}
in addition to the conditions derived in section \ref{sec:Lnot0}.  
Note that the penultimate line above implies that the 
Lie derivative of the background three-form field vanishes, $\mathcal{L}_KH=0$; the second line can also be written as $\mathcal{L}_KL=0$; the third line is equivalent to $K$ being Killing. It also follows from the above that the norm of $K$ is constant, ${\partial_m|K|^2=0}$. Hence for $K\neq0 $ the triplet $(K,J,\Omega)$ defines on $\mathcal{M}_6$ an $SU(2)$ structure. 

Normalizing $|K|^2=2$ the $SU(2)$ structure can be given equivalently 
as the triplet $(K,j,\omega)$, where \cite{Gauntlett:2003cy, bovy}:
\eq{\omega=-\frac{i}{2}K^*\lrcorner\Omega~,}
is a complex (2,0) form and 
\eq{
j=J-\frac{i}{2}K\wedge K^*
~,}
is a real (1,1) form obeying the compatibility conditions
\eq{
\omega\wedge j=0~,~~~\omega\wedge\omega^*=2j\wedge j
~.}
If desired, the last line in (\ref{11cond}) may be expressed in terms of 
$SU(2)$ torsion classes.

Note again that choosing $E_m=0$ in the first line of (\ref{11cond}) implies that the manifold is conformally balanced and the scalar $\xi_H$ defined in (\ref{const}) vanishes.

\subsection{(1,2) supersymmetry}\label{sec:12}

Imposing $E_m=0$ and consequently the conformally balanced condition $W_4=d\phi$, eqs.~(\ref{11cond}) leave the constant $c$ unconstrained. The spinor $\zeta$ in (\ref{z}) can then be written as 
$\zeta=c\eta+c'\eta'$ where $\eta'=\frac1{\sqrt{2}|K|}\,\gamma^m K_m \eta^c$ 
is a unimodular spinor orthogonal to $\eta$ and $c'=\sqrt{2} |K|$\,.
Therefore, $\zeta$ is parameterized by two independent 
parameters $c$ and $c'$, hence the theory is (at least) (1,2)-supersymmetric. 
In that case it can be seen  that the equations in (\ref{11cond})  reduce to:
\eq{\spl{\label{11condsimpl}
\partial K&=0\;,\\
\mathcal{L}_K\phi&=0\;,\\
\nabla_{(m}K_{n)}&=0\;,\\
d\left(e^{-2\phi}K\wedge J \right)
&=0~,}}
where we have set $L=e^{-2\phi}$. In deriving the equations above we have 
used eq.~(\ref{b4}) and have taken into account that
\eq{
W_3^{(2,1)}=\partial J-J\wedge\partial\phi
~,}
which follows from (\ref{torsionclassesb}) and the exactness of $W_4$.

\subsection{Examples}\label{sec:examples}

There are many examples of six-dimensional manifolds obeying the necessary and 
sufficient conditions (\ref{tc}), (\ref{tc0}). One obvious class is that of Calabi-Yau manifolds; in this case all torsion classes vanish. 
A compact non-Ricci-flat, non-conformally-balanced example with (1,1) supersymmetry is that of the round $\mathbb{S}^3\times\mathbb{S}^3$. We will also provide  non-compact examples based on the Iwasawa manifold (with (1,1) supersymmetry), as well as certain T$^2$ bundles over non-compact K3's (with (1,2) supersymmetry). As we will see, however, our rigid-supersymmetric theory cannot be defined on the round six-sphere, at least for $\theta^r\neq0$. The latter conclusion is in accordance with the representation-theoretic classification of Euclidean supersymmetries \cite{nahm}, as pointed out recently in \cite{Kehagias:2012fh}. Note however that the classification of \cite{nahm} only considers ``space-times'' whose isometries are simple Lie groups. In particular it does not cover the cases without any isometries, such as e.g. Calabi-Yau manifolds.

\subsubsection*{The round $\mathbb{S}^3\times\mathbb{S}^3$}

This example is based on an $SU(3)$ structure given 
in \cite{Gutowski:2002bc}. We identify $\mathbb{S}^3\times\mathbb{S}^3$ with 
the group manifold SU(2)$\times$SU(2). The orthonormal frame is given by the 
SU(2)-invariant one-forms $e^a$, $f^a$, $a=1,2,3$, satisying
\eq{\label{s3}
d e^a=\frac{1}{2}\varepsilon_{abc}e^b\wedge e^c
~~~~d f^a=\frac{1}{2}\varepsilon_{abc}f^b\wedge f^c
~.}
The $SU(3)$ structure is given by:
\eq{\spl{\label{su3s3}
J&=e^1\wedge e^2-e^4\wedge e^5+e^3\wedge e^6\\
\Omega&=(e^1+i e^2)\wedge(e^4-i e^5)\wedge(e^3+i e^6)
~,}}
so that a basis of (1,0)-forms is given by 
$(e^1+i e^2)$, $(e^4-i e^5)$, $(e^3+i e^6)$. 
The metric associated  with the structure (\ref{su3s3}) is that of the 
round  $\mathbb{S}^3\times\mathbb{S}^3$:
\eq{
ds^2=\sum_{a=1}^3\left(e^a\otimes e^a+f^a\otimes f^a\right)
~.}
Moreover we obtain $W_1$, $W_2=0$ (hence $\mathcal{M}_6$ is hermitian), and
\eq{\spl{\label{ts3}
W_3&=\frac{1}{2}(e^1\wedge e^2+f^1\wedge f^2)\wedge (f^3-e^3)\\
W_4&=-\mathrm{Re}W_5=\frac{1}{2}(e^3+f^3)\\
W_5&=e^{\frac{3\pi i}{4}}~\!\frac{~\!e^3+if^3}{\sqrt{2}}
~,}}
as can be seen by computing $dJ$, $d\Omega$ taking (\ref{s3}) into account and 
comparing with eq.~(\ref{torsionclassesb}). 
The $H$-field can be computed from (\ref{tcbb}) and (\ref{ts3}):
\eq{
H=-(e^1\wedge e^2\wedge e^3+f^1\wedge f^2\wedge f^3)
~,}
which is closed, $dH=0$ (equivalently $\partial\bar{\partial}J=0$, hence $\mathcal{M}_6$ is strong K\"{a}hler with torsion). Furthermore, 
setting $L=0$, $E_m=0$, the round $\mathbb{S}^3\times\mathbb{S}^3$ satisfies the necessary and sufficient conditions (\ref{tc0}) and thus provides a consistent  
background for a rigid (1,1)-supersymmetric theory. Note in particular that 
this example is not conformally balanced: indeed it follows from (\ref{ts3}) that $W_4$ is not closed and therefore not exact.

\subsubsection*{The noncompact Iwasawa manifold}

The Iwasawa manifold can be viewed as a compact quotient $G/\Gamma$ where $G$ is the three-dimensional complex Heisenberg group defined as 
\eq{G=\left\{
\left( \begin{array}{ccc}
 ~\!1~\! & ~\!z^1\! &  -z^3 \\
0 & 1& z^2 \\
0 & 0 &1 \end{array} \right)~\!,~~z^1,z^2,z^3\in\mathbb{C}
\right\}
~,
}
with group product given by ordinary matrix multiplication, 
and $\Gamma$ acts as
\eq{
\Gamma : \left( \begin{array}{c}
z^1 \\
z^2 \\
z^3 \end{array} \right) \longrightarrow  {\bf M}\cdot\left( \begin{array}{c}
z^1 \\
z^2 \\
z^3 \end{array} \right)+{\bf m}
~,}
where 
\eq{
{\bf m}=\left( \begin{array}{c}
m^1 \\
m^2 \\
m^3 \end{array} \right)\in\mathbb{Z}^3\oplus i\mathbb{Z}^3
~,
 ~~{\bf M}=\left( \begin{array}{ccc}
 ~\!1~\! & ~\!0~\! &  ~\!0~\! \\
0 & 1& 0 \\
0 &-m^2 &1 \end{array} \right)
~.}
By `noncompact Iwasawa' we mean the group manifold $G$ (which is homeomorphic to $\mathbb{C}^3$), i.e. before taking the $\Gamma$ quotient.

The left invariant forms on $G$ are given by 
$dz^1$, $dz^2$, $dz^3+z^1dz^2$. 
We can choose the orthonormal frame of one-forms $e^a$, $a=1,\dots, 6$ so that 
\eq{
e^1+i e^2=dz^1~, ~~ e^3+i e^4=dz^2~, ~~e^5+i e^6=dz^3+z^1dz^2~. 
}
It follows 
that:
\eq{\spl{\label{iwas}
d e^a&=0~,~~~a=1,\dots 4\\
d e^5&=e^1\wedge e^3-e^2\wedge e^4\\
d e^6&=e^1\wedge e^4+e^2\wedge e^3
~.}}
The $SU(3)$ structure is given by
\eq{\spl{
J&=e^{2\phi}\left(e^1\wedge e^2+e^3\wedge e^4\right)+e^5\wedge e^6\\
\Omega&=e^{2\phi}(e^1+i e^2)\wedge(e^3+i e^4)\wedge(e^5+i e^6)
~,}}
so that the $G$ left-invariant forms given above are (1,0); $\phi$ is 
a function of $z^1$, $z^2$ but does not depend on $z^3$. 
The metric associated with (\ref{iwas}) reads
\eq{
ds^2=\sum_{a=1}^4e^{2\phi}\left(e^a\otimes e^a\right)
+\sum_{a=5}^6e^a\otimes e^a
~.}
We obtain $W_1$, $W_2=0$, and
\eq{\spl{W_3&=e^1\wedge e^3\wedge e^6-e^2\wedge e^4\wedge e^6-e^1\wedge e^4\wedge e^5-e^2\wedge e^3\wedge e^5\\
&~~~+\frac{1}{2}d\phi\wedge\left[
e^{2\phi}\left(e^1\wedge e^2+e^3\wedge e^4\right)-e^5\wedge e^6
\right]\\
W_4&=-\mathrm{Re}W_5=d\phi
~,}}
as can be seen by computing $dJ$, $d\Omega$ taking (\ref{iwas}) into account and 
comparing with eq.~(\ref{torsionclassesb}). Moreover, the $H$-field can be computed from (\ref{tcbb}) and (\ref{ts3}):
\eq{
H=-e^2\wedge e^4\wedge e^5+e^1\wedge e^3\wedge e^5+e^2\wedge e^3\wedge e^6+e^1\wedge e^4\wedge e^6
-\star\left(
d\phi\wedge e^5\wedge e^6
\right)
~.}
Imposing closure, $dH=0$, is equivalent to 
\eq{
\sum_{i=1}^2\frac{\partial}{\partial z^i}\frac{\partial}{\partial \bar{z}^i}e^{2\phi}+1=0
~,}
which can only be satisfied for the noncompact Iwasawa.

This example is conformally balanced since $W_4$ is exact. 
Hence by setting the background field $E_m$ to zero 
the noncompact Iwasawa satisfies the 
necessary and sufficient conditions (\ref{tc}) and moreover 
we obtain a (1,1)-supersymmetric theory. Note that although the vector dual to the (1,0)-form $K=e^5+ie^6$ is Killing (corresponding to isometries along the $z^3$ direction), $K$ is not $\partial$-closed: $\partial K= dz^1\wedge dz^2$. Hence the conditions (\ref{11condsimpl}) for (1,2) supersymmetry are not satisfied.

\subsubsection*{T$^2$ bundles over noncompact K3}

This example is based on the work of 
\cite{gold} (see also \cite{dasg}). In that reference it was shown that  
six-dimensional manifolds with $SU(3)$ structure 
can be constructed as T$^2$ fibrations over 
K3 surfaces. The metric on the total space is given by 
\beal
ds^2=e^{2\phi}ds^2(\mathrm{K3})+(dx+\alpha)^2+(dy+\beta)^2~,
\label{mk3}
\end{align}
where $ds^2(\mathrm{K3})$ is the Ricci-flat metric on K3, $\phi$ is a  function on K3,  and $\alpha$, $\beta$ are local one-forms on K3 which satisfy $d\alpha=\omega_P$, $d\beta=\omega_Q$ with
\beal
\frac{[\omega_P]}{2\pi},~\frac{[\omega_Q]}{2\pi}\in H^{2}(\mathrm{K3},\mathbb{Z})\cap  H^{1,1}(\mathrm{K3})~.
\label{quantization}
\end{align}
The complex $SU(3)$ structure on the total space is 
given by
\eq{\spl{
J&=e^{2\phi}j+(dx+\alpha)\wedge(dy+\beta)\\
\Omega&=e^{2\phi} \omega\wedge\left[(dx+\alpha)+i(dy+\beta)  \right] ~,
\label{totj}
}}
where $\omega$, $j$ are the holomorphic $(2,0)$ form,  
the K\"{a}hler form on the K3 base\footnote{If the map $\pi : \mathcal{M}_6\mapsto \mathcal{M}_6/ T^2 \simeq \mathrm{K3}$ 
defines the fibration, we can extend $\omega$, $j$, $\alpha$, $\beta$ 
from K3 to the total space $\mathcal{M}_6$ by using $\pi^*$.}
 respectively. In particular $\omega$, $j$ are both closed. 
Moreover we will assume that $\omega_P$, $\omega_Q$ are anti self-dual $\star\omega_{P,Q}=-\omega_{P,Q}$, so that\footnote{
Consider  the decomposition 
\eq{
H^2(\mathcal{M}_4,\mathbb{R})=H^+(\mathcal{M}_4,\mathbb{R})\oplus H^-(\mathcal{M}_4,\mathbb{R})
~,}
into self-dual and anti self-dual two-forms. For  $\mathcal{M}_4$
a compact K3 surface, $H^+(\mathcal{M}_4,\mathbb{R})$ is three-dimensional and is generated by $\mathrm{Re}\omega$, $\mathrm{Im}\omega$ and $j$; $H^-(\mathcal{M}_4,\mathbb{R})$ is the 19-dimensional vector space of real (1,1)-forms which are orthogonal to $j$ in the sense of (\ref{orthj}).}
\eq{\label{orthj}
\omega_P\wedge j=\omega_Q\wedge j=0
~.}
By computing $dJ$, $d\Omega$ and comparing with (\ref{torsionclassesb}) it can then be seen that the torsion classes are given by
\eq{\spl{
W_3&=\frac12 d\phi\wedge\left[
e^{\phi}j-(dx+\alpha)\wedge(dy+\beta)
\right]-(dx+\alpha)\wedge\omega_Q+(dy+\beta)\wedge\omega_P\\
W_4&=-\mathrm{Re}W_5=d\phi
~.}}
From the above and (\ref{hstar}) we find that the $H$ field is given by 
\eq{
\star H=-\omega_P\wedge(dy+\beta)+\omega_Q\wedge(dx+\alpha)
+d\phi\wedge(dx+\alpha)\wedge(dy+\beta)~.
}
The closure of $H$ can then be seen to be equivalent to
\eq{
\nabla_{\mathrm{K3}}^2e^{2\phi}+\left|\omega_P\right|^2
+\left|\omega_Q\right|^2=0~,
}
where $\nabla_{\mathrm{K3}}^2$ is the (unwarped) Laplacian on K3, 
and we have defined $\left|\omega\right|^2=\omega^*_{mn}\omega^{mn}/2$. 
Similarly to the previous example, the above equation can only be solved 
for noncompact K3's.

Since $W_4$ is exact we can set the background field $E_m$ to zero, 
hence this example satisfies the necessary and sufficient conditions 
(\ref{tc}) and possesses (1,1) supersymmetry. 
Moreover let us define the one form $K$ by
\eq{
K=dz+(\alpha+i\beta)
~,}
where $z=x+iy$, 
which is (1,0) with respect to the complex structure associated with (\ref{totj}). Clearly the vector dual to $K$ is Killing  for the metric in (\ref{mk3}). Furthermore $\mathcal{L}_K\phi=0$ since $\phi$ only depends on the coordinates of K3 and not on the T$^2$ coordinate $z$. We also have 
$dK=\omega_P+i\omega_Q\in H^{1,1}(\mathrm{K3})$, hence $\partial K=(dK)^{2,0}=0$. Finally from (\ref{totj}) we find
$$
d\left(e^{-2\phi}K\wedge J\right)=d\left(K\wedge j\right)
=(\omega_P+i\omega_Q)\wedge j=0~.
$$
Therefore all the conditions in (\ref{11condsimpl}) are satisfied and we obtain a rigid (1,2)-super\-symmetric theory.

\subsubsection*{The round sphere $\mathbb{S}^6$}

Whether or not the six-dimensional sphere admits a complex structure is still an open question. It is known however that the six-dimensional sphere does not admit an orthogonal complex structure, i.e. one that obeys eq.~(\ref{osb})  with respect to the standard `round' metric \cite{lebr}. In other words the round sphere $\mathbb{S}^6$ is not a hermitian manifold. This implies that $\mathbb{S}^6$ violates the first of the necessary and sufficient conditions in (\ref{tc}) therefore our rigid-supersymmetric theory cannot be defined on $\mathbb{S}^6$, at least for $\theta^r\neq0$.


\section{Euclidian $6D$ SYM and localization}
\label{sec:local}


In this section we shall set the parameters $\theta^r$ and all other background fields to zero,
\eq{\label{2}
L^\circ=B^\circ=V^\circ=E^\circ=\theta^r=0
~,}
except for the metric of the curved six-dimensional space $\mathcal{M}_6$. Then the analysis of section \ref{sec:kse2} implies that all torsion classes vanish and $\mathcal{M}_6$ is a Calabi-Yau threefold\footnote{We will use the term `Calabi-Yau threefold' to refer to any six-dimensional manifold whose holonomy is a subset of $SU(3)$, including T$^6$ and T$^2\times$K3.}. Equivalently, this can be seen by the fact that imposing  (\ref{2}), the first Killing spinor equation (\ref{KS}) reduces to the condition that the spinor $\epsilon$ is covariantly constant, while eq.~(\ref{KS0}) is automatically satisfied. Moreover, as follows from the analysis of section \ref{sec:11}, this example possesses (1,1) supersymmetry since $\theta^r=0$. The upshot is that Euclidean rigid (1,1)-supersymmetric nonabelian YM theory can be consistently defined on CY threefolds.

\subsection{Localization}

In order to apply the localization procedure, the theory must be invariant under the action of a fermionic operator $Q$ which is nilpotent, $Q^2=0$ (or more generally squares to a symmetry of the theory). Deforming the action by a $Q$-exact term,
\eq{
S\longrightarrow S+ \frac{1}{e^2}Q\cdot U
~,}
leaves invariant the expectation values of $Q$-closed operators. Hence we may take the limit of zero coupling constant, $e^2\rightarrow 0$, upon which the theory localizes to the set $\Sigma$ of critical points  of $Q\cdot U$ \cite{wittentop}. 
In this limit the path integral can be performed by restricting $S$ to $\Sigma$ and computing a one-loop determinant describing the fluctuations normal to $\Sigma$. This procedure has been carried out in detail in e.g. \cite{pest} for the case of SYM on the round $\mathbb{S}^4$.

The action (\ref{Lag_vec}) is only invariant on-shell.  
In order to construct the fermionic operator $Q$ of the previous paragraph 
we introduce auxiliary scalar bosonic fields $K^r$. The operator $Q$ is then defined by its action on the fields:
\bea
Q\cdot A_m{}^r &=& \tilde{\nu}^r \gamma_m \eta
\;,\nonumber\\
Q\cdot \lambda^r &=& \frac14\,\gamma^{mn}\eta\,F_{mn}{}^r
-\frac{1}{2}\,K^r\eta
\nonumber\\
Q\cdot K &=&\tilde{\eta}\gamma^m D_m\nu^r  
\;,
\label{Q}
\eea
where $\eta$ is a commuting unimodular Weyl spinor, as in section \ref{sec:kse2}, which is covariantly constant:
\eq{
\nabla_m\eta=0
~.}
Note that by virtue of the fact that $\eta$ is commuting $Q$ defined in (\ref{Q}) is indeed fermionic, i.e. transforms commuting to anticommuting fields and vice-versa.

It can be checked that the operator $Q$ defined in (\ref{Q}) is nilpotent off-shell, i.e. $Q^2=0$ on all fields without using the equations of motion. 
Furthermore it can be checked that $Q$ leaves the following Lagrangian invariant 
up to a total derivative: 
\eq{\label{lagoff}
{\cal L}_{\mathrm{off}} =
\frac14\,F_{mn}{}^r F^{mn\,r}-2 \tilde{\nu}^r \gamma^m D_m\,\lambda^r
+\frac{1}{2}K^rK^r
\;,
}
where ${\cal L}_{\mathrm{off}}$ reduces to pure on-shell SYM upon eliminating the auxiliary fields by their equations of motion. Explicitly we have:
\eq{
Q\cdot\mathcal{L}_{\mathrm{off}}=\nabla_m\left[
(\tilde{\eta}\gamma^m\nu^r)K^r
\right]
~.}
The sign difference between the $F^2$ term in (\ref{Lag_vec}) and that in (\ref{lagoff}) above comes from the fact that the spinor parameter $\eta$ in (\ref{Q}) is commuting whereas $\epsilon$ in (\ref{susy_vector}) is anticommuting.

From (\ref{lagoff}) we see that for convergence of the path integral the fields $A^r_m$, $K^r$ should be real. On the other hand the 
$Q$-transformations of $A^r_m$, $K^r$ given in (\ref{Q}) are complex. This does not affect the localization argument  \cite{pest}: the action should be thought of as an analytic functional in the space of complexified fields, with the path integral 
understood as integration over the real subspace thereof.

We now deform the Lagrangian by a $Q$-exact term:
\eq{
{\cal L}_{\mathrm{off}}\longrightarrow{\cal L}_{\mathrm{off}}+\frac{1}{e^2}Q\cdot U
~,}
where $U$ is given by
\eq{
U=\lambda^\dagger Q\cdot\lambda
~.}
A straightforward computation using (\ref{Q}) then gives
\eq{\spl{\label{qv}
Q\cdot U&=
\left(Q\cdot\lambda^r\right)^{\dagger}(Q\cdot\lambda^r)\\
&=\frac{1}{8}\left(
F_{mn}{}^rF^{mn~\!r}-\frac{1}{4}\varepsilon^{m_1\dots m_6}
F_{m_1m_2}{}^rF_{m_3m_4}{}^rJ_{m_5m_6}
\right)
+\frac{1}{4}K^r K^r 
~,}}
where we have taken the definition (\ref{o}) into account. Note that since 
we are in the case where $\mathcal{M}_6$ is a CY threefold and the spinor 
$\eta$ is covariantly constant, $J_{mn}$ is a K\"{a}hler form and 
$J_m{}^{n}$ is a complex structure.

According to the localization procedure, the theory will localize, in the $e^2\rightarrow 0$ limit, to the set $\Sigma$ of critical points of $Q\cdot U$. To determine $\Sigma$ we note that the right-hand side of (\ref{qv}) can be expressed as a sum of squares:
\eq{\label{squares1}
Q\cdot U=
\frac{1}{16} 
\left(
F^{mn~\!r}-\frac{1}{4}\varepsilon^{mnm_1\dots m_4}
F_{m_1m_2}{}^rJ_{m_3m_4}-\frac{1}{2}J^{mn}F_{pq}{}^rJ^{pq}
\right)^2
+\frac14\left(K^r\right)^2
~,
}
where we have taken into account (\ref{osa}) and the identity:
\eq{
\varepsilon^{mnm_1\dots m_4}
J_{m_1m_2}J_{m_3m_4}=-8J^{mn}
~,}
which can be proven e.g. by fierzing, taking into account the definition 
(\ref{o}) of $J$ as a spinor bilinear.

To bring (\ref{squares1}) to a more familiar form, it is useful to project on the (2,0), (0,2) and (1,1) parts (with respect to the 
complex structure $J_m{}^n$) using the projectors defined in 
(\ref{projectors}). Explicitly we expand:
\eq{
F_{mn}{}^{r}=F_{mn}^{(2,0)~\!r}+F_{mn}^{(0,2)~\!r}+\tilde{F}_{mn}^{(1,1)~\!r}+\frac{1}{6}F_0^rJ_{mn}
~,}
where
\eq{\spl{
F_0^r&={F}_{mn}{}^{r}J^{mn}\\
F_{mn}^{(2,0)~\!r}&=\left(\Pi^+\right)_m{}^p\left(\Pi^+\right)_n{}^qF_{pq}{}^r\\
F_{mn}^{(0,2)~\!r}&=\left(\Pi^-\right)_m{}^p\left(\Pi^-\right)_n{}^qF_{pq}{}^r
=\left(F_{mn}^{(2,0)~\!r}\right)^*\\
\tilde{F}_{mn}^{(1,1)~\!r}&=
2\left(\Pi^+\right)_{[m}{}^p\left(\Pi^-\right)_{n]}{}^qF_{pq}{}^r-\frac{1}{6}F_0^rJ_{mn}~,
}}
so that $\tilde{F}_{mn}^{(1,1)~\!r}$ is (1,1) and traceless: 
$\tilde{F}_{mn}^{(1,1)~\!r}J^{mn}=0$. 
With these definitions, eq.~(\ref{squares1}) can be rewritten as\footnote{
More generally, for complex auxiliary fields $K^r$ we find
\eq{\label{squares2}
Q\cdot U=
\frac{1}{2}\left|F_{mn}^{(2,0)~\!r}\right|^2
+\frac{1}{4}\left(\mathrm{Im}K^r-\frac{1}{2}F_0^r\right)^2
+\frac{1}{4}\left(\mathrm{Re}K^r\right)^2
~.
}
As we have already mentioned, for convergence of the euclidean 
path integral we must take the auxiliary fields to be real.
}
\eq{\label{squares3}
Q\cdot U=
\frac{1}{2}\left|F_{mn}^{(2,0)~\!r}\right|^2+\frac{1}{16}\left(F_0^r\right)^2
+\frac{1}{4}\left(K^r\right)^2
~.
}
Hence the set of critical points of $Q\cdot U$ is given by:
\eq{\spl{
F_{mn}^{(2,0)~\!r}&=F_{mn}^{(0,2)~\!r}=0\;,\\
F_0^r&=0\;,\\
K^r&=0
~,\label{sb}}}
while the (1,1)-traceless component of the field strength remains unconstrained. 
These equations are of course familiar from heterotic compactifications on Calabi-Yau manifolds: The first line of (\ref{sb}) defines a holomorphic gauge bundle over $\mathcal{M}_6$; the second line is the Donaldson-Uhlenbeck-Yau equation defining a stable bundle.

In conclusion: the path integral of the SYM theory (\ref{lagoff}) defined on a CY threefold $\mathcal{M}_6$ localizes on stable holomorphic bundles over $\mathcal{M}_6$. Note however that we have restricted our discussion to essentially classical considerations. A quantum-mechanical treatment taking into account the path-integral measure will not be attempted here.

\section{Conclusions}
\label{sec:conclusions}

In this paper, we have analyzed the conditions for rigid supersymmetry in six dimensions
and translated them as a set of necessary and sufficient conditions on the geometry of the six-dimensional manifold. 
In particular, the background three-form flux $H_{kmn}^\circ$ is given in terms of the torsion classes.
Given the existence of Killing spinors on these background structures, we have constructed the
explicit Lagrangians and field equations for the interactions of
non-abelian vector- and tensor-multiplets on curved space.
We have worked out a number of explicit examples, including CY threefolds, 
the round $\mathbb{S}^3\times \mathbb{S}^3$, as well as noncompact examples based on 
the Iwasawa manifold and $T^2$ bundles over noncompact K3's. 
Finally, we have applied the localization procedure to pure SYM theory
defined on a CY threefold ${\cal M}_6$ and shown that 
the path integral localizes on stable holomorphic 
bundles over ${\cal M}_6$.


A number of interesting questions remains to be further analyzed.
We have fully analyzed the conditions for rigid backgrounds allowing for
the minimal amount of supersymmetries, i.e.\ $N=(1,0)$ and $N=(1,1)$ in our notation,
and briefly discussed the consequences of $N=(1,2)$. 
The existence of more Killing spinors will impose further constraints on the
background fields and geometry, just as in the four-dimensional case~\cite{Festuccia:2011ws}.
With increasing number of supercharges, the structure of 
the supersymmetry algebra becomes more and more constraining.
Let us note that the algebra displayed in~(\ref{susy_algebra}), and~(\ref{susy_algebra_ten}), 
is soft in the sense that the closure involves dependence on fields, including the background 
ones. Taking the flat space limit in which the dynamical fields and the background 2-form 
potential vanishes, yields the rigid superalgebra
\bea
\{\tilde Q_\alpha,Q_\beta\} &=& (C\gamma^\mu)_{\alpha\beta} P_\mu\ ,
\eea
which is the Euclidean Poincar\'e superalgebra in $6D$ \cite{Lukierski:1985yb}. If we set only the dynamical fields equal to zero, the interpretation of the resulting superalgebra depends on the details of the background fields, such as the Killing spinors and and isometries they may support.  
In this case, a superalgebra in lower dimensions which also contains internal symmetry generators may be expected. In particular, the analogs of the Poincar\'e, AdS and conformal superalgebras for space-times in arbitrary dimensions and signatures have been systematically constructed in \cite{DAuria:2000ec}. 

Another interesting open question concerns possible anomalies of our model.
In Minkowski space-time and with all background fields set to zero, the model reduces to that 
of~\cite{Bergshoeff:1996qm}. In that model there are gauge anomalies due to the coupling of gauge fields to chiral fermions. For certain gauge groups these anomalies can be cancelled by Green-Schwarz mechanism involving the addition of a one-loop counterterm of the  form $\hbar B\wedge F\wedge F$.
Supersymmetrization of the model in the presence of this counterterm \cite{Howe:1998ts} leads to ${\hbar}$-corrections to the Yang-Mills field equation, which however is anomalous, as its divergence does not vanish, as expected. This also implies an anomaly term in the closure of the superalgebra. Nonetheless, as shown in \cite{Howe:1998ts}, introducing a constant $\alpha'$ in front of the YM interactions in the tensor multiplet field equations, and then  interchanging the role of $\alpha'$ with ${\hbar}$ leads to a dual model such that  (a)  it is anomaly free upon treating the resulting ${\hbar}$-terms as one-loop counterterm corrections, and (b) the tensor multiplet equations are free, while they arise as source in the field equations of the Yang-Mills multiplet.  While the Euclideanization of $6D$ space-time is not expected to upset these results, the consequences of retaining nontrivial background fields from the context of the anomalies  are not clear, and remains an i!
 nteresting open problem.
Nonetheless, a direct application of the Noether procedure  in the dual formulation that uses the Killing spinor equations is expected to yield a formulation of these models in curved background. It is also worth noting that the tensor-YM system \`a la~\cite{Bergshoeff:1996qm} as well as its dual formulation, both in $6D$ Minkowski spacetime, were also obtained from different global limits of the (on-shell) heterotic supergravity compactified on K3,  \cite{Duff:1997me}.

It should be straightforward to extend the localization results presented here to non-Calabi-Yau manifolds. This would entail modifying the analysis of section 
\ref{sec:local} to include non-vanishing background fields besides the metric. 
We should stress however that our treatment of localization was essentailly limited to classical considerations. A fully quantum-mechanical treatment should include the consideration of the path-integral measure and potential anomalies.

\subsection*{Acknowledgments}
We would like to thank K.\ Becker and C.\ Pope for helpful discussions.
This work is supported in part by the F\'ed\'eration de Physique Andr\'e Marie Amp\`ere.
E.S.\ thanks Lyon University where part of this work was done for hospitality. 
The research of E.S.\ was supported in part by NSF grant PHY-0906222.

\bigskip
\bigskip
\bigskip

\begin{appendix}

\section{6D conventions}
\label{app:6D}

In this section we list some useful relations and 
explain in more detail our spinor conventions for general even-dimensional Euclidean spaces of dimension $2k$. 

The charge conjugation matrix obeys:
\eq{\label{c}
C^{\mathrm{Tr}}=(-)^{\frac{1}{2}k(k+1)}C~;~~C^*=(-)^{\frac{1}{2}k(k+1)}C^{-1}~;~~
\gamma_m^{\mathrm{Tr}}=(-)^k C^{-1}\gamma_m C
~.
}
The complex conjugate $\eta^c$ of a spinor $\eta$ is given by:
\eq{
\eta^c:=C\eta^*
~,}
form which it follows that:
\eq{
(\eta^c)^c=(-)^{\frac12 k(k+1)}\eta
~.}
Covariantly-transforming spinor bilinears must be of the 
form $(\widetilde{\psi}\gamma_{m_1\dots m_p}\chi)$, 
where in any dimension we define:
\bea
\widetilde{\psi}:=\psi^{\mathrm{Tr}}C^{-1}
~.
\label{tilde}
\eea
One can also show the following useful identity:
\eq{
\gamma_{m_1\dots m_p}^*=(-)^{kp}C^{-1}\gamma_{m_1\dots m_p}C
~.}

The case of six-dimensional Euclidean space is obtained by specializing to 
 $k=3$. The chirality matrix is given by
\bea
\gamma_{m_1 \dots m_6} &=& -i \, \varepsilon_{m_1 \dots m_6}\,\gamma^7
\;.
\eea
Since the chiral irreducible representation of $Spin(6)$ is complex, 
 a Weyl spinor $\eta$ and its complex conjugate $\eta^c$ have opposite chirality. We also have:
\eq{
\tilde{\eta^c}=\eta^{\dagger}
~.}
Moreover we find:
\eq{\label{bilins}
(\widetilde{\psi}_\pm\gamma_{m_1\dots m_{2l}}\chi_\pm)=0=(\widetilde{\psi}_\pm\gamma_{m_1\dots m_{2l+1}}\chi_\mp)~, 
}
where the subscript $\pm$ denotes the chirality. 
Another useful relation is:
\eq{\label{bilinstr}
(\widetilde{\psi}\gamma_{m_1\dots m_p}\chi)=
\sigma(-)^{\frac12 p(p+1)}(\widetilde{\chi}\gamma_{m_1\dots m_p}\psi)
~,}
where $\sigma=+1$, $\sigma=-1$ for commuting, anticommuting spinors $\chi$, $\psi$.
As for the spinor fields encountered in our models, the supersymmetry 
parameters ($\epsilon$, $\tilde\zeta$) and the gauginos ($\lambda$, $\tilde\nu$) 
have chirality $+$ whereas the tensorinos ($\chi$, $\tilde\psi$) have chirality $-$.

\mathon
\section{$SU(3)$-structure identities}\label{app:b}
\mathoff

Here we summarize several relations which are useful in analyzing the 
Killing spinor equations. We follow ref.~\cite{Lust:2004ig} which the reader may 
consult for further details. 
\beal
0&=(\Pi^+)_m{}^n\gamma_n\eta^c\nn\\
\gamma_{mn}\eta&=iJ_{mn}\eta+\frac{1}{2}\Omega_{mnp}\gamma^p\eta^c\nn\\
\gamma_{mnp}\eta^c&=-3iJ_{[mn} \gamma_{p]}\eta^c-\Omega^*_{mnp}\eta~.
\end{align}
The above together with tensor decomposition (\ref{hexp}) give
\beal
\gamma^{mnp}H_{mnp}\eta&=H^{(0)*}\eta^c+3iH^{(0,1)}_t\gamma^t\eta\nn\\
H_{mpq}
\gamma^{pq}\eta&=\Big(H^{(0)*}g_{mt}
+3\Omega_t{}^{pq}\tilde{H}_{mpq}^{(1,2)}
-\frac{3i}{2}\Omega_{mt}{}^pH_p^{(0,1)}
\Big)\gamma^t\eta^c+6i(H_m^{(1,0)}+H_m^{(0,1)}  )\eta\nn\\
H_{pqr}
\gamma^{pqr}\gamma_m\eta&=3\Omega_t{}^{pq}\tilde{H}_{mpq}^{(1,2)}\gamma^t\eta^c
+6iH_m^{(1,0)}
~.
\label{a1}
\end{align}
In the above we have taken into account the identity
\be{}
\Omega_{[t}{}^{pq}\tilde{H}_{m]pq}^{(1,2)}=0
~,
\ee
which can easily be shown by direct calculation.  

The following identity is useful in deriving the formul\ae{} 
in section \ref{sec:12}:
\eq{\label{b4}
(V\lrcorner W)\wedge J=iV\wedge W
~,}
where $V$ is an arbitrary (1,0)-form and $W$ is any three-form obeying $W\wedge J=0$.

\end{appendix}


\providecommand{\href}[2]{#2}\begingroup\raggedright\endgroup

\end{document}